\documentclass{sig-alternate}
\usepackage{verbatim}
\usepackage{pstricks,pst-plot,pst-node,pst-grad,pstricks-add}
\usepackage{pst-3dplot}
\usepackage{multirow}
\usepackage{xspace}
\usepackage{subfigure}
\usepackage{balance}
\usepackage{url}

\makeatletter
\newif\if@restonecol
\makeatother

\usepackage[vlined,ruled,linesnumbered]{algorithm2e}

\newcommand{\alg}{DAGGER\xspace}
\newtheorem{remark}{Remark}
\newtheorem{property}{Property}

\begin{document}

\title{\alg: A Scalable Index for Reachability Queries\\ in 
Large Dynamic Graphs}

\numberofauthors{1} 
%
\author{
\alignauthor
Hilmi Yildirim$^1$, Vineet Chaoji$^2$, and Mohammed J. Zaki$^3$
~\thanks{This work was done when the first author was at RPI and it was supported in part 
by NSF Grants EMT-0829835 and NIH Grant 1R01EB0080161-01A1.}\\
\affaddr{$^1$Google, Pittsburgh PA USA}\\
\affaddr{$^2$Amazon, Bangalore, India}\\
\affaddr{$^3$Rensselaer Polytechnic Institute, Troy NY USA}\\
\email{hilmi@google.com, vchaoji@amazon.com, zaki@cs.rpi.edu}
}

\maketitle
\begin{abstract}
With the ubiquity of large-scale graph data in a variety of application
domains, querying them effectively is a challenge.
In particular, reachability queries are becoming increasingly
important, especially for containment, subsumption, and
connectivity checks. Whereas many methods have been proposed for
static graph reachability, many real-world graphs are constantly
evolving, which calls for dynamic indexing. In this paper, we
present a fully dynamic reachability index over dynamic graphs.
Our method, called \alg, is a light-weight index based on
interval labeling, that scales to million node graphs and beyond.
Our extensive experimental evaluation on real-world and synthetic
graphs confirms its effectiveness over baseline methods.
\end{abstract}





\section{Introduction}

Graph-based representation of data has become predominant with
the emergence of large-scale interlinked networks, such as 
social networks, biological networks, the World Wide Web and semantic RDF
(Resource Description Framework) graphs. For instance, Facebook
has 750 million users with and average of 130 friends per user.
This implies that the Facebook social graph has 750 million
nodes, with approximately 49 billion edges.
Similarly, RDF graphs with over a billion triples are quite
common these days.
 
Most of the above real world networks undergo update operations.
These updates include addition and deletion of edges or nodes.
In social networks such as
Twitter and Facebook, it is not surprising to see a dynamically
changing graph structure as new connections emerge or existing
ones disappear. Web graph undergoes frequent updates with
new links between pages. Wikipedia is a representative example
wherein links are added as new content pages are generated, and
deleted as corrections are made to the content pages.

The scale of these datasets has renewed interest in 
graph indexing and querying algorithms. Answering reachability
queries in graphs is one such area.  Given a directed graph
$G=(V,E)$, where $V$ is the set of vertices and $E$ is the set of
directed edges, a \textit{reachability query} asks if there
exists a path $p$ from a source node $u$ to a target node $v$ in
the directed graph $G$.  If such a path exists, we say that $u$
can reach $v$ (or $v$ is reachable from $u$), and denote it as $u
\rightsquigarrow v$. If $u$ cannot reach $v$, we denote it as $u
\not\rightsquigarrow v$. The reachability query itself is denoted
as $u \stackrel{?}{\rightsquigarrow} v$.  Traditional
applications for reachability assumed that the graph
was static.

The new emerging applications such as social network analysis,
semantic networks, and so on, however, call for reachability
queries on dynamic graphs.  For example, social networks 
rely extensively on updates in
order to recommend new connections to existing users 
(e.g., via the `People You May Know' feature in Facebook). Within dynamic RDF
graphs, reachability queries help determine the relationships
among pairs of entities.

Both the scale and the dynamic nature of these graphs call for
highly scalable indexing schemes that can accommodate graph
update operations like node/edge insertion and deletion.
Recomputing the entire index
structure for every update is obviously computationally
prohibitive. Moreover, for online systems that receive a steady
volume of queries, recomputing the index would result in system
down-time during index updation.  As a result, a reachability
index that can accommodate dynamic updates to the underlying
graph structure is desired. Despite this need, the dynamic
reachability problem has received scant attention. This is primarily due to the complex nature
of the problem -- a single edge addition or deletion can
potentially affect the reachability of all pairs of nodes in the
graph. Moreover, most of the static indexes cannot be directly
generalized to the dynamic case. This is because these indexes
trade-off the computationally intensive preprocessing/index
construction stage to minimize the index size and querying time.
For dynamic graphs, the efficiency of the update operations is
another aspect which needs to be optimized.  However, the costly
index construction typically precludes fast updates.  It is
interesting to note that a simple approach consisting of
depth-first search (DFS) can handle graph updates in $O(1)$ time
and queries in $O(n+m)$ time\footnote{For sparse graphs $m=O(n)$ so that query time is $O(n)$ for most
large real-world graphs.}, where $m$ is the number of edges.
Any dynamic index will be effective only
if it can amortize the update costs over very many reachability
queries.

\begin{figure}[!h]
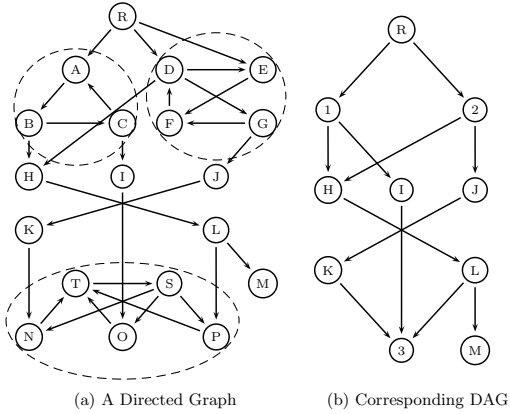

	\vspace{-0.1in}
\scriptsize
\centerline{
\scalebox{0.7}{
\hspace{-0.2in}
\subfigure[A Directed Graph]{\label{fig:sample:a}
 \psset{xunit=0.35in, yunit=0.4in}
\pspicture[](-2,-1)(5.5,6.5)
 \psset{arrows=->,arrowscale=1,nodesep=2pt}
 \rput(1,6){\circlenode{R}{R}}
 \rput(0,5){\circlenode{A}{A}}
 \rput(-1,4){\circlenode{B}{B}}
 \rput(1,4){\circlenode{C}{C}}
 \rput(2,5){\circlenode{D}{D}}
 \rput(4,5){\circlenode{E}{E}}
 \rput(2,4){\circlenode{F}{F}}
 \rput(4,4){\circlenode{G}{G}}
  \rput(-1,3){\circlenode{H}{H}}
  \rput(1,3){\circlenode{I}{I}}
  \rput(3,3){\circlenode{J}{J}}
  \rput(-1,2){\circlenode{K}{K}}
  \rput(3,2){\circlenode{L}{L}}
 \rput(-1,0){\circlenode{N}{N}}
 \rput(1,0){\circlenode{O}{O}}
 \rput(3,0){\circlenode{P}{P}}
 \rput(0,1){\circlenode{T}{T}}
 \rput(2,1){\circlenode{S}{S}}
  \rput(4,1){\circlenode{M}{M}}
  \ncline{R}{A}
  \ncline{R}{D}
  \ncline{R}{E}
  \ncline{A}{B}
  \ncline{B}{C}
  \ncline{C}{A}
  \ncline{D}{E}
  \ncline{F}{D}
  \ncline{D}{G}
  \ncline{E}{F}
  \ncline{G}{F}
  \ncline{B}{H}
  \ncline{C}{I}
  \ncline{G}{J}
  \ncline{D}{H}
  \ncline{H}{L}
  \ncline{J}{K}
  \ncline{L}{M}
  \ncline{K}{N} 
  \ncline{I}{O}
  \ncline{L}{P}
  \ncline{N}{T}
  \ncline{O}{T}
  \ncline{P}{T}
  \ncline{T}{S}
  \ncline{S}{N}
  \ncline{S}{O}
  \ncline{S}{P}
\psellipse[linewidth=0.02,dimen=outer,linestyle=dashed](0.0,4.3)(1.33,1.1)
\psellipse[linewidth=0.02,dimen=outer,linestyle=dashed](3.0,4.5)(1.5,1.2)
\psellipse[linewidth=0.02,dimen=outer,linestyle=dashed](1,0.3)(2.5,1.1)
\endpspicture
}
\hspace{-0.15in}
\subfigure[Corresponding DAG]{\label{fig:sample:b}
 \psset{xunit=0.55in, yunit=0.60in}
 \psset{arrows=->,arrowscale=1,nodesep=2pt}
\pspicture[](0.0,-0.5)(2.5,4.5)
 \rput(1,4){\circlenode{R}{R}}
 \rput(0,3){\circlenode{A}{1}}
  \rput(2,3){\circlenode{B}{2}}
  \rput(0,2){\circlenode{C}{H}}
  \rput(1,2){\circlenode{D}{I}}
  \rput(2,2){\circlenode{E}{J}}
  \rput(0,1){\circlenode{F}{K}}
  \rput(2,1){\circlenode{G}{L}}
  \rput(1,0){\circlenode{H}{3}}
  \rput(2,0){\circlenode{I}{M}}
  \ncline{R}{A}
  \ncline{R}{B}
  \ncline{A}{C}
  \ncline{A}{D}
  \ncline{B}{C}
  \ncline{B}{E}
  \ncline{C}{G}
  \ncline{D}{H}
  \ncline{E}{F}
  \ncline{F}{H}
  \ncline{G}{H}
  \ncline{G}{I}
  \endpspicture
  }}
}
	\vspace{-0.1in}
\caption{\small Sample input graph and its DAG}
\label{fig:sample}
\end{figure}

Let us consider the example graph in Figure~\ref{fig:sample:a}. It
is worth noting at the outset that the nodes $A$, $B$ and $C$
have identical reachability because they form a strongly connected
component (SCC). Coalescing such components (shown within dashed
ovals) into a single node yields a directed acyclic graph (DAG)
called the condensation graph, as depicted in
Figure~\ref{fig:sample:b}. Letters are used for the initial
graph labels and numbers for the SCC nodes (except for single
node SCCs, which remain as letters).  For instance, the
SCC $\{A,B,C\}$ is represented as node 1, $\{D,E,F,G\}$ as 2, and
$\{N,O,P,S,T\}$ as 3.  In the static setting, all reachability
queries can be answered over the DAG.  However for dynamic
graphs, maintaining the DAG structure imposes additional
overhead. First consider inter-component edges.  For example, the
deletion of the edge $(H,L)$, affects the reachability of nodes
$R$, $1$, $2$, and $H$. Adding the edge $(C,J)$ only impacts the
reachability of node $1$, which now can reach nodes $J$, and $K$.
On the other hand, adding the edge $(N,B)$ creates a new SCC
composed of $1$, $H$, $I$, $L$ and $3$. Therefore the
corresponding DAG has to be updated by merging these nodes into a
new SCC labeled 4 (not shown).  Now consider, an intra-component
edge; deleting $(D,G)$ splits SCC $2$ into two components
$(D,E,F)$ and $G$. Furthermore, this update causes the nodes $D$,
$E$ and $F$ to lose their reachability to $J$ and $K$.  These
examples clearly show that local changes to the graph can have
widespread impact in terms of the reachability.

In this paper, we propose a scalable, light-weight reachability
index for dynamic graphs called \alg (an anagram of the bold
letters in {\bf D}ynamic {\bf G}raph {\bf REA}chability, with an
extra `G'), 
which has linear (in the
order of the graph) index size and index construction time,
and reasonably fast query and update times. Some updates
can be handled in constant time, however, updates can take
linear time in the worst case. 
In particular, we make the following contributions:
\begin{list}{\labelitemi}{\leftmargin=1em}
	\item \alg uses dynamic interval labels based on multiple
		random traversals of the SCC DAG.
	\vspace{-0.1in}
	\item \alg supports common update operations, namely node
		and edge insertions and deletions.  Updates made to the
		input graph are mapped onto updates on the corresponding
		DAG over the SCC nodes.  Additions (deletions) that do
		not merge (split) SCCs are accommodated in constant time.
		For updates that merge or split SCCs, the DAG is
		appropriately updated. 
  %
	\vspace{-0.1in}
	\item Whereas many of the previous approaches have been
		tested on relatively small graphs (with up to 400k
		nodes), we perform a comprehensive set of experiments
		over graphs with millions of nodes and edges. 
		To our knowledge, this is also the first work that
		experimentally evaluates updates for all four
		operations. We
		explicitly study the tradeoff between indexing and searching
		in the presence of dynamic updates.
\end{list}

Our experimental evaluation confirms that \alg is a scalable,
light-weight, and dynamic reachability index that outperforms
existing approaches, especially as the ratio of queries to
updates increases.

\section{Related Work}
\label{sec:survey}

Many algorithms have been proposed for answering 
reachability queries on
static graphs. They can be broadly categorized into 
two main groups -- interval
labeling~\cite{agrawal,tribl,wang,jin,chen2008}, and 2HOP
labeling~\cite{cohen,schenkel_hopi,cheng,cheng2,he2}. 
The {\em interval
labeling} approaches use either the min-post
labeling~\cite{agrawal}, or pre-post
labeling~\cite{tribl,chen2008,jin}, on a spanning subtree of
the DAG of the original graph.  {\em Pre-post labeling} assigns
$L_u = [s_u,e_u]$ to each node $u$ where $s_u$ and $e_u$ are the
pre-order and post-order ranks of node $u$ in a DFS traversal of
the DAG, starting from the root(s), with the rank being
incremented each time we enter a node or back-track from a node.
In contrast, in a min-post labeling, $e_u$ is the post-order rank
of $u$, and $s_u$ is the minimum rank of any node under $u$. 

In {\it 2HOP indexing} \cite{cohen, schenkel, schenkel_hopi, cheng, cheng2,
he2} each node determines a set of intermediate nodes it
can reach, and a set of intermediate nodes which can reach it.
The query between $u$ and $v$ returns success if the intersection
of the successor set of $u$ and predecessor set of $v$ is not
empty.  Hybrid approaches that combine 2HOP and interval labeling
also exist~\cite{he2}.

\smallskip \noindent{\bf Dynamic Indexing Methods:} 
While the above techniques focus on reachability in static
graphs, not much attention has been paid to practical algorithms
for the dynamic case.  

The interval label based Optimal Tree Cover (Opt-TC)~\cite{agrawal},  while
primarily a static index, was also one of the first works to
address incremental maintenance of the index.  Opt-TC first
creates interval labels for a spanning tree of the DAG. However,
for a non-tree edge (i.e., an edge that is not part of the
spanning tree), say between $u$ and $v$, $u$ inherits all the
intervals associated with node $v$.  Testing reachability is
equivalent to deciding whether the interval of the source node
subsumes the interval of the target node. Since selecting the
optimal tree cover requires pre-computing the transitive closure,
this method is computationally infeasible for large graphs. Certain operations
such as addition and deletion of non-tree edges involve updating the
intervals of all the predecessors. Due to the significant 
overheads associated with incremental maintenance of the
optimal cover, Opt-TC is practically infeasible for large graphs.

In~\cite{bramandiaJ}, the authors propose a technique for 
incremental maintenance of the 2-HOP labeling in the presence of
graph updates -- namely, addition and deletion of nodes and
edges. The delete operation in a typical 2-HOP labeling is
expensive, since it requires the updation of the successor and
predecessor labels of the deleted node/edge. To overcome this drawback, the
authors propose alternate 2-HOP labelings ({\em node separable 2-HOP labeling})
that rely on heuristics based on cut vertex and minimum graph bisection. On the
down side, the cover constructed using these heuristics has much larger index
size as compared to the static case.

For reachability in static graphs, HOPI~\cite{schenkel_hopi}
builds a 2-HOP cover on partitions of the original graph that can
fit in memory; these are then merged by adding label entries for
links between partitions. Subsequent work~\cite{schenkel} by the
same authors extends HOPI to allow incremental maintenance of the
HOPI index.  Like~\cite{bramandiaJ}, the delete operation is the
most compute intensive for HOPI index as well.
 
Within the theory community, several works have focused on the
theoretical aspects of reachability in the presence of updates to
the graph~\cite{roditty, henzinger, king-jcss02, demetrescu2000}.
Note that most of these algorithms maintain the
entire transitive closure in memory and relies on a fast matrix
multiplication procedure for the update operation. 
It is important to note that these methods are designed for dynamic
transitive closure, and as such are not suitable for reachability
queries in very large graphs, due to their quadratic space
requirements. 
Recently, several of these theoretical methods for 
dynamic transitive closure were
experimentally analyzed in \cite{krommidas2008}.
However,
scalability remains an issue as they require quadratic space.

\section{The \alg Approach}

Static graphs are easier to index due to the fact that index
construction is performed only once, and therefore there is
tolerance for super-linear construction complexity.  In a dynamic
setting, where queries and graph updates are intermixed, a small
change in the graph structure might end up altering the labels of
many nodes. For example, removal of a bridge edge or an
articulation node may alter the
reachability status of $O(n^2)$ pairs. Given that upper bound,
each operation in a dynamic index should be performed in
sublinear time with respect to the graph order/size, otherwise
one can naively reconstruct a linear-time index 
after each operation instead of using the dynamic index.

\alg is an interval labeling based reachability index for dynamic
graphs. Like other interval labeling methods it works on the DAG
structure of the input graph, and thus in the dynamic case, it
has to handle update operations on the strongly connected
components of the graph. \alg is thus also a method to actively
maintain the corresponding DAG. \alg assigns multiple (randomized)
interval labels to every DAG node. It supports the four basic
update operations on the input graph, which are insertion and
deletion of an edge, and insertion and deletion of a node along
with its incident (incoming/outgoing) edges. 

\newcommand{\vshearbox}[3]{\scalebox{#2}[0.866025]{\rotatebox{90}%
{\scalebox{-0.57735}[1.73205]{\rotatebox{95}{\scalebox{#1}[-1.1547]{#3}}}}}}

\begin{figure}[!ht]
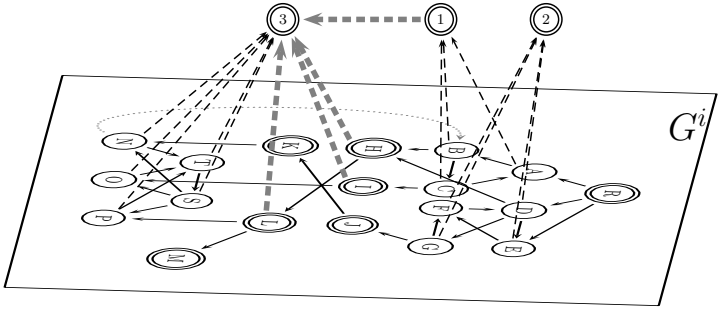

	\vspace{-0.1in}
\small
\hspace{0.1in}
\scalebox{0.7}{%
\pspicture[](-3.5,-4.5)(3.5,4.5)
 \psset{arrows=->,arrowscale=1,nodesep=2pt}
\rput(8,1.5){\huge $G^i$}
\vshearbox{1.0}{0.90}{
\rotatebox{90}{
 \pspolygon[](-4,-3)(3,-3)(3,5)(-4,5)
 \psset{xunit=0.25in, yunit=0.4in}
 \rput(-1,4){\circlenode[doubleline=true]{R}{R}}
 \rput(-2,3){\circlenode{A}{A}}
 \rput(-3,2){\circlenode{B}{B}}
 \rput(-1,2){\circlenode{C}{C}}
 \rput(0,3){\circlenode{D}{D}}
 \rput(2,3){\circlenode{E}{E}}
 \rput(0,2){\circlenode{F}{F}}
 \rput(2,2){\circlenode{G}{G}}
  \rput(-3,1){\circlenode[doubleline=true]{H}{H}}
  \rput(-1,1){\circlenode[doubleline=true]{I}{I}}
  \rput(1,1){\circlenode[doubleline=true]{J}{J}}
  \rput(-3,0){\circlenode[doubleline=true]{K}{K}}
  \rput(1,0){\circlenode[doubleline=true]{L}{L}}
 \rput(-3,-2){\circlenode{N}{N}}
 \rput(-1,-2){\circlenode{O}{O}}
 \rput(1,-2){\circlenode{P}{P}}
 \rput(-2,-1){\circlenode{T}{T}}
 \rput(0,-1){\circlenode{S}{S}}
  \rput(3,-1){\circlenode[doubleline=true]{M}{M}}
  \ncline{R}{A}
  \ncline{R}{D}
  \ncline{R}{E}
  \ncline{A}{B}
  \ncline{B}{C}
  \ncline{C}{A}
  \ncline{D}{E}
  \ncline{F}{D}
  \ncline{D}{G}
  \ncline{E}{F}
  \ncline{G}{F}
  \ncline{B}{H}
  \ncline{C}{I}
  \ncline{G}{J}
  \ncline{D}{H}
  \ncline{H}{L}
  \ncline{J}{K}
  \ncline{L}{M}
  \ncline{K}{N} 
  \ncline{I}{O}
  \ncline{L}{P}
  \ncline{N}{T}
  \ncline{O}{T}
  \ncline{P}{T}
  \ncline{T}{S}
  \ncline{S}{N}
  \ncline{S}{O}
  \ncline{S}{P}
  \nccurve[linecolor=gray,linestyle=dotted,linewidth=1pt,dotsep=1pt,angleA=-135,angleB=180]{N}{B}
}
}
 \rput(3,3.5){\circlenode[doubleline=true]{1}{1}}
 \ncline[linestyle=dashed]{A}{1}
 \ncline[linestyle=dashed]{B}{1}
 \ncline[linestyle=dashed]{C}{1}
 \rput(5,3.5){\circlenode[doubleline=true]{2}{2}}
 \ncline[linestyle=dashed]{D}{2}
 \ncline[linestyle=dashed]{E}{2}
 \ncline[linestyle=dashed]{F}{2}
 \ncline[linestyle=dashed]{G}{2}
 \rput(0,3.5){\circlenode[doubleline=true]{3}{3}}
 \ncline[linestyle=dashed]{S}{3}
 \ncline[linestyle=dashed]{T}{3}
 \ncline[linestyle=dashed]{O}{3}
 \ncline[linestyle=dashed]{N}{3}
 \ncline[linestyle=dashed]{P}{3}
 \ncline[linestyle=dashed,linewidth=3pt,linecolor=gray]{1}{3}
 \ncline[linestyle=dashed,linewidth=3pt,linecolor=gray]{H}{3}
 \ncline[linestyle=dashed,linewidth=3pt,linecolor=gray]{I}{3}
 \ncline[linestyle=dashed,linewidth=3pt,linecolor=gray]{L}{3}
\endpspicture
}
\vspace{-0.75in}

\caption{\small \alg graph: i) Initial input graph $G^i=(V^i,E^i)$ is
shown on the plane (solid black edges).  The SCC components
($V^d$) are shown as double-circled nodes.  Components consisting
of single nodes are shown on the plane, whereas larger components
are shown above the plane. DAG edges $E^d$ are not shown for
clarity.  Containment edges ($E^c$) are shown as black dashed
arrows.  ii) Insertion of the (dotted gray) edge $(N,B)$ in
$G^i$, merges five SCCs, namely $\{1,H,I,L,3\}$, with $3$ as the
new representative. The thick gray dashed edges are the new
containment edges. }
\vspace{-0.1in}
\label{fig:SCC}
\end{figure}

\subsection{\alg Graph}

\alg maintains a consolidated, layered graph structure to
represent the input graph, as well as the DAG structure, and the
containment relationships between nodes and components.
Formally, the \alg graph is given as $G = (V,E)$, where $V = V^i
\cup V^d \cup V^e$, and $E = E^i \cup E^d \cup E^c$, where these
node/edge sets are defined as follows:

\vspace{-0.1in}
\hspace{-0.9cm}
\begin{list}{\labelitemi}{\leftmargin=1em}

\item {\bf Input Graph ($G^i$):} The input graph is denoted as
	$G^i = (V^i, E^i)$, with the node set $V^i$ and edge set
	$E^i$. Any update operation is first applied to $G^i$ which
	later impacts the other constituents of $G$.

\item {\bf DAG ($G^d$):} Due to the split/merge operations on the
	SCCs resulting from graph updates, SCCs are of two types: {\em
	current} or {\em expired}. An expired SCC is one that has
	been subsumed (merged) into another SCC at some point.  The
	condensation graph of the input graph is a DAG $G^d = (V^d,
	E^d)$, where each node in $V^d$ represents a current SCC of
	the updated input graph, and $E^d$ consists of edges between
	SCCs implied by the updated input graph.  That is, $E^d =
	\{(s,t) | s,t \in V^d, \text{ \it and there exists an edge }
	(u,v) \in E^i \text{ \it such that } s=S(u) \text{ \it and }
	t=S(v)\}$, where $S(u)$ denotes the current SCC corresponding
	to the input node $u \in V^i$.  
	Note that \alg also keeps track of the number of such edges
	between SCCs, i.e., there might be different pairs of input
	edges $(u_i,v_i)$, with $S(u_i)=s$ and $S(v_i)=t$. \alg
	stores this multiplicity information on the edge itself.
	Also $size(s)$ denotes the
	number of nodes comprising the SCC $s$.  Note that the set
	$V^e$ refers to the expired SCCs, whereas $V^d$ constitutes the
	{\em current} DAG nodes after any update operation.
	We refer to the current set of nodes/edges
	in $G^d$ as the {\em DAG nodes/edges}. 

\item {\bf Containment edges ($E^c$):} These refer to the
	subsumption relationships between input nodes and SCCs, or
	between SCCs.  Thus for a node $u \in V^i$, the containment
	edge $(u,t)$ implies that node $u$ belongs to the SCC $t$,
	whereas for a node $s\in V^e$, the containment edge $(s,t)$
	implies that all nodes in the expired SCC $s$ are contained
	in SCC $t$ (which may be expired or current).  Containment
	edges constitute a union-find data structure where the leaf
	nodes (with no children/subsuming nodes) represent the set of
	current SCCs, i.e., the DAG nodes $V^d$. The current
	SCC $S(u)$ for any node $u \in V^i$ can thus be found by
	tracing a path from $u$ to a leaf component node, via
	containment edges.

\end{list}

Figure~\ref{fig:SCC} shows the \alg graph corresponding to the
example graph in Figure~\ref{fig:sample}. The input graph $G^i$ is
shown on the bottom plane, whereas the SCC nodes are placed
higher.  The input edges $E^i$ are shown as solid lines, whereas
the input nodes $V^i$ are single- or double-circled, and labeled
with letters. The initial set of containment edges are shown as
black dashed arrows (ignore the gray thick dashed arrows for
now).  The initial set of DAG nodes, the current SCCs, are shown
double circled. A SCC node containing a single input node, e.g.,
$R$, is shown double-circled on the bottom plane, whereas a SCC
node with $size > 1$ is labeled with a number, and shown above
the plane, e.g., SCC $2$ represents nodes $D$, $E$, $F$ and $G$,
which is also shown via the containment edges from those nodes to
$2$.  The current set of DAG nodes is thus $V^d =
\{1,2,3,R,H,I,J,K,L,M\}$. It is important to note that this is
precisely how \alg avoids duplicating the DAG structure, i.e.,
whereas we have used $V^i$ and $V^d$ to denote the input and DAG
nodes, for clarity. In our implementation, the fact whether a
node is an input node, or a SCC node is conveniently represented
using appropriate node labels. This is important, since real
world graph are large and sparse, and \alg can thus avoid
duplicating large parts of the graph which are DAG-like (e.g., in
the extreme case, if the input graph is a DAG, then $G^i$ and
$G^d$ would be identical, and thus \alg can cut down the space by
half).  Note that Figure~\ref{fig:SCC} does not show the DAG edges $E^d$ to
avoid clutter; these edges would be precisely those shown in
Figure~\ref{fig:sample:b}.  Finally, Figure~\ref{fig:SCC} also 
shows what
happens due to an update operation, namely the addition of the
(dotted gray) edge $(N,B)$.  This causes SCC $1$ to be merged
into SCC $3$ (as described later), along with nodes $H, I, L$.
These changes are reflected via the dashed, thick gray
containment edges. After this update, the set of current SCCs is
$V^d = \{2,3,R,J,K,M\}$, and the set of expired ones is $V^e =
\{1\}$ (note that we do not include the single node SCCs in $V^e$
as another space saving optimization).  

Throughout the paper, we will use the letters $u$ and $v$ to
refer to nodes in $V^i$ (the input nodes), 
and $s$ and $t$ to refer to nodes in $V^d$ (the DAG nodes).

\subsection{Interval Labeling}

\alg maintains the DAG structure corresponding to the input graph
updates. It thus maintains labels only for the SCC nodes. 
A {\em labeler} $L: V^d \to \{L^i = [b^i,e^i]\}_{i=1}^k$, with
$b^i,e^i \in \mathbb{N}$, is a function that assigns a
$k$-dimensional interval to every SCC node. We refer to the label
of node $s$ as $L_s$, while $L_s^i$ is the $i^{th}$ dimension of
the label. We refer to the beginning of the interval $L_s^i$ as
$b^i_s$ and the ending as $e^i_s$.

\alg is a light-weight reachability index that uses relaxed
interval labeling, which makes it suitable for indexing dynamic
graphs. The only invariant it maintains is that if a node $s$
reaches $t$, $L_s$ has to subsume $L_t$. Equivalently, if $L_s^i$
does not subsume $L_t^i$ for any $i$, $s$ definitely does not
reach $t$.  After each update on the input graph, \alg updates
labels based on the changes to $G$.

\begin{property}{No False Negative: }
\label{property}
 If $s \rightsquigarrow t$ then $L_t \subset L_s$. Equivalently, if 
$L_t \not\subset L_s$ then $s \not\rightsquigarrow t$.
\end{property}

\alg uses relaxed interval labeling as follows: Assume we know
the labels of each child $t$ of node $s$. The tightest interval,
$L_s^i = [b_s^i, e_s^i]$ (for each dimension $i \in [1,k]$]), that we can assign to $s$ would be to start from the
minimum of $b^i_t$ (i.e., $b_s^i = \min_t \{b^i_t\}$), 
and to end at the maximum of $e^i_t+1$ (i.e., $e_s^i = \min_t
\{e^i_t+1\}$), over
all children $t$. 
However, we do not use the tightest possible
intervals.  This is because we want flexibility in assigning
labels due to split/merge operation on the SCCs.  For instance,
when a SCC $s$ is split into multiple components due to an edge
deletion, the interval of $s$ has to be shared between the new
components which might not be possible if we use very tight
intervals.  Instead, our scheme maintains a gap of at least
$size(s)$ in $L_s$. 

Querying in \alg exploits the property~\ref{property}. Given 
query $u \rightsquigarrow v$, we lookup their components
$s=S(u)$ and $t=S(v)$. If $L_s$ does not subsume $L_t$, we 
conclude that $s \not\rightsquigarrow t$. Otherwise, the search
continues from the children of $s$ recursively until we find a 
path to $t$, or the search can be pruned earlier. 

\subsection{Supported Operations}

\alg supports the following update operations that constitute a
fully dynamic setting for directed graphs.

\vspace{-0.1in}
\hspace{-0.9cm}
\begin{list}{\labelitemi}{\leftmargin=1em}

\item {\textbf InsertEdge($u$,$v$):} Adding an edge between two
	nodes that are in the same SCC does not change the
	reachability of any pairs of nodes. Similarly if $s=S(u)$ and
	$t=S(v)$, and there exist a DAG edge $(s,t)$, it will have no
	effect on the reachability. On the other hand, if $(s,t)$
	does not exist, then all the descendants of $t$ will become
	reachable from the all the ancestors of $s$. In \alg's
	interval labeling, this can be accommodated by enlarging the
	intervals of the ancestors of $s$ so that they contain the
	interval of $t$.  The worst case occurs when $t$ also reaches
	$s$, in which case at least two components have to be merged,
	which alters the DAG structure and the corresponding
	labeling.

\item {\textbf DeleteEdge($u$,$v$)} If $u$ and $v$ are in
	different components $s$ and $t$, respectively, and there are
	at least two edges between $s$ and $t$, the removal of $(u,v)$
	has no effect on the DAG structure and labeling. If $(u,v)$
	is the only edge between $s$ and $t$, the DAG edge
	$(s,t)$ has to be removed, and the index has to be updated.
	Lastly, if the nodes are in the same component, the edge
	removal might split the components into many smaller
	components which can lead to a costly update operation on the
	labels. This is especially true for large 
	real world graphs that usually contain giant
	strongly connected components.

\item {\textbf InsertNode($u$,$E_u$):} We support node addition,
	along with its set of outgoing and incoming edges. 
	\alg first adds the node, and then handles the edges via a
	series of edge insertions.

\item {\textbf DeleteNode($u$):} When we delete a node we
	also have to delete all incoming and outgoing edges, which
	are handled as a series of edge deletions. However, in this
	case, it is much more likely that a component splits and
	some components become disconnected.
\end{list}

The biggest challenge for an efficient reachability index on a
dynamic graph is maintaining the strongly connected components
efficiently, especially given the fact that almost all of the
existing methods are designed to work on DAGs.  Below, 
we describe the details of \alg's interval labeling
(Section~\ref{sec:algdetails}), and
we show how these interval labels are
maintained over the DAG in response to the above graph update
operations (Section~\ref{sec:maint}).

\section{\alg Construction}
\label {sec:algdetails}

\subsection{Initial Graph Construction}
Given an input graph $G^i$, \alg uses Tarjan's algorithm~\cite{tarjan} to find the strongly connected components.
For each SCC that has more than one node, 
we create a SCC node $s$ in $V^d$, and we connect the constituent
input nodes to $s$ via containment edges in $E^c$. 
If a node $u$ in $G^i$ is 
itself a component, we do not create a SCC node for it. 
The black colored (solid/dashed) nodes/edges in
Figure~\ref{fig:SCC} show the initial \alg graph for our example
graph in Figure~\ref{fig:sample}.
In this graph, the SCCs $1$, $2$ and $3$ are created during the
construction.  We also compute the DAG edges 
$E^d = \{ (1,H), (1,I), (2,H), (2,J) \}$, 
which are not shown in Figure~\ref{fig:SCC} for readability purposes. 
As an space-saving optimization, we do not add
DAG edges between SCC nodes comprising single input nodes.
As noted previously, single node SCCs are not added to $V^d$, 
which helps reduce the space by at most a factor of two, with the
limit achieved when $G^i$ is already a DAG.

\newcommand{\threelabelnode}[4]{${#2}\stackrel{{\red #4}}{#1}${#3}}
\newcommand{\twolabelnode}[3]{${#2}~{#1}$~{#3}}

\begin{figure}[!ht]
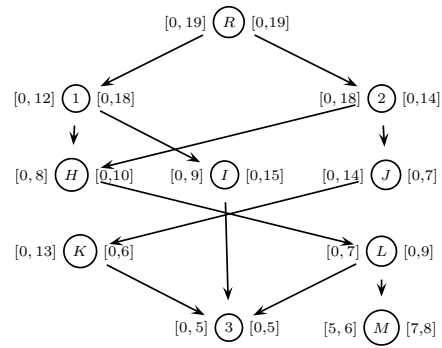

	\scriptsize
\scalebox{0.8}{%
\centerline{
\hspace{-0.5in}
 \psset{xunit=1in, yunit=0.5in}
\pspicture[](-1,-0.5)(2.5,4.5)
 \psset{arrows=->,arrowscale=1.5,nodesep=6pt}
 \rput(1,4){\twolabelnode{\circlenode{R}{R}}{[0,19]}{[0,19]}}
 \rput(0,3){\twolabelnode{\circlenode{A}{1}}{[0,12]}{[0,18]}}
  \rput(2,3){\twolabelnode{\circlenode{B}{2}}{[0,18]}{[0,14]}}
  \rput(0,2){\twolabelnode{\circlenode{C}{H}}{[0,8]}{[0,10]}}
  \rput(1,2){\twolabelnode{\circlenode{D}{I}}{[0,9]}{[0,15]}}
  \rput(2,2){\twolabelnode{\circlenode{E}{J}}{[0,14]}{[0,7]}}
  \rput(0,1){\twolabelnode{\circlenode{F}{K}}{[0,13]}{[0,6]}}
  \rput(2,1){\twolabelnode{\circlenode{G}{L}}{[0,7]}{[0,9]}}
  \rput(1,0){\twolabelnode{\circlenode{H}{3}}{[0,5]}{[0,5]}}
  \rput(2,0){\twolabelnode{\circlenode{I}{M}}{[5,6]}{[7,8]}}
  \ncline{R}{A}
  \ncline{R}{B}
  \ncline{A}{C}
  \ncline{A}{D}
  \ncline{B}{E}
  \ncline{C}{G}
  \ncline{E}{F}
  \ncline{G}{H}
  \ncline{G}{I}
  \ncline{B}{C}
  \ncline{D}{H}
  \ncline{F}{H}
\endpspicture
}}
\vspace{-0.1in}
\caption{\small Two valid initial labelings}
\vspace{-0.1in}
\label{fig:initial}
\end{figure}
\subsection{Initial Label Assignment}

We use a modified min-post labeling scheme to label the initial
SCC nodes. \alg performs multiple ($k$ of them) randomized
traversals on the DAG $G^d$,  to assign multiple intervals to the
nodes, following the basic approach outlined in GRAIL
index~\cite{grail}. For each traversal, labeling starts from the root nodes of
the DAG, and a label of a node is assigned after all of its
children are labeled, i.e., we use post-order traversals. During
the recursive traversal \alg keeps a counter {\em ctr}, 
which is incremented by the
size of the SCC node $s$ when exiting the node $s$.
Figure~\ref{fig:initial} shows an example interval labeling of the
DAG $G^d$, using $k=2$ (i.e., with two interval labels per node).  Each node has the first label on its
left, and second label on its right. In the first traversal, the
nodes are visited in left to right order whereas the order is
reversed in the second traversal (ordering is non-randomized just
for illustration).  The first traversal ($i=1$) arrives as node
$3$ for the first time via the path $(R,1,H,L,3)$ and thus
assigns the interval $[0,5]$ to $L_3^1$ since the SCC size of
$3$, $size(3)$, is 5.  Then it backtracks to $L$ and visits its
next child $M$, assigning it the interval $[5,6]$. Note that
$b_M^1=5$, since that is the value of $ctr$ when $M$ is visited,
and $e_M^1 = 5+1=6$, since $size(M)=1$.  After labeling the nodes
$L$ and $H$, the traversal visits node $I$, whose child $3$ has
already visited.  In this case, $I$ gets an interval that starts
from the minimum $b^i_t$ of all its children $t$, thus the
interval is $L_I^1 = [0,9]$, since $ctr=8$ upon entry and
$size(I)=1$.  The second traversal visits and labels the nodes in
the order $(R,2,J,K,3,H,L,M,1,I)$, which leads to the second set
of intervals shown in Figure~\ref{fig:initial}.

\alg makes $k$ such random traversals to
form $k$ intervals for each node, encapsulated as $L_s = L_s^1,
L_s^2, \cdots, L_s^d$ for each DAG node $s$.  
Note that having multiple
traversals helps  avoid false positives. For instance, looking at
only the
first interval $L_2^1=[0,18]$ subsumes $L_1^1=[0,12]$ 
although SCC node $2$ does
not reach $1$. However, looking at the second interval,
$L_1^2=[0,18]$ subsumes
$L_2^2=[0,14]$. Thus, with both intervals considered
simultaneously, neither the hyper-rectangle $L_1$ nor 
$L_2$ subsume each other. 
It is worth noting that \alg's labeling leaves enough
gap in the label of each node so that when a component node is
split, its interval can be shared among the new nodes. 
Details of how labels are updated will be given later.

\subsection{Component Lookup}

Containment edges ($E^c$) and current/expired 
SCC nodes ($V^d \cup V^e$) comprise a
union-find~\cite{unionfind} data structure within \alg. Such a
union-find structure is more efficient than directly maintaining
a lookup table for a node $x$ and its SCC $S(x)$, especially for
merging components. 
For instance, if we are merge
$k$ SCCs, each of which represents $b$ nodes (on average), 
via union-find we can merge them in $O(k)$ time, whereas it would
take $O(bk)$ time via a lookup table.

Finding the SCC node $S(u)$ that represents input node $u$ is
straightforward. The process starts from $u$ and follows the
containment edges as long as it can.  The node that does not have
a containment edge is the corresponding SCC node.  We apply two
optimizations to provide faster lookup performance. First, when
we are merging several components we always attach other
components to the largest component. Secondly, whenever we lookup
a node $u$, we update all the containment edges of the component
nodes on the path from $u$ to $S(u)$. This is known as path
compression~\cite{unionfind}.  These two optimizations provide
$O(\alpha(n))$ amortized lookup time where $\alpha(n)$ is the
inverse Ackermann function. Ackermann function is such a
quickly-growing function that its inverse is smaller than 5 for
any practical value of $n$.  Therefore component lookup is
essentially constant time.

\section{\alg: Dynamic Maintenance}
\label{sec:maint}
For each of the update operations on the input graph, we first
show how we update the \alg graph $G$. Once the graph is updated,
we explain how to update the labels for the DAG nodes.

\subsection{Edge Insertion}
\label{sec:edgeinsertion}

\subsubsection{SCC Maintenance}
Given an edge $e=(u,v)$ to be inserted in $G^i$, we first 
locate their corresponding components $s=S(u)$ and $t=S(v)$, and
check whether they are equal.  If $u$ and $v$ are in the same
component, no further action has to be taken.  Otherwise, we
check whether the insertion of $e$ merges some of the existing
components.

\begin{remark}
An edge insertion merges at least two SCC nodes if and only if
$s$ is already reachable from $t$, i.e., if $t \rightsquigarrow
s$.

Proof: If there is already a path from $t$ to $s$, insertion of
$e=(s,t)$ completes the cycle creating a new component which
contains all the nodes en route from $t$ to $s$ (note: there may
be multiple such cycles).
\label{rem:edgeinsertion} 
\end{remark}

We can thus check if we need to merge some components via the
reachability query
$t \stackrel{?}{\rightsquigarrow}s$.  For example, consider the
insertion of the dotted gray edge $(N,B)$ to $G^i$ in
Figure~\ref{fig:SCC}. The corresponding SCC nodes are $S(N)=3$ and
$S(B)=1$.  We query $1\stackrel{?}{\rightsquigarrow}3$, which
returns a positive answer. Thus,  all the nodes involved in paths
that start from $1$ and end at $3$ will be the members of the new
component. 

In general, there are two cases to consider when inserting an
edge.  The first case is that $t \not\rightsquigarrow s$, and
$(s,t) \not\in E^d$. In this case, the only change in the \alg
graph $G$ is the addition of $(s,t)$ into $E^d$.  If $(s,t) \in
E^d$ no change is required at all.  The second case is when $t
\rightsquigarrow s$, creating at least one cycle.  To find all
the nodes en route from $t$ to $s$, \alg performs a recursive
search starting from $t$. The algorithm adds a node $w$ to the
result list if any of its children lead to a path to $s$.  In
other words if $w\not\rightsquigarrow s$, $w$ should not be in
the list. Here too, we take advantage of \alg labels, since if
$L_t \not\subset L_s$, then $w\not\rightsquigarrow s$. In that
case we do not recurse into $w$. While finding the nodes en route
from $1$ to $3$ in Figure~\ref{fig:initial}, the search starts from
$1$ and proceeds with $H$ and $L$ but it does recurse from $M$
since $L_3 \not\subset L_M$.  Even if there was a big subgraph
under $M$, the pruning would prevent us from visiting those nodes
unnecessarily.  On the other hand, since $L$ also has an edge to
$3$, $L$ is included in the result list. This, in turn, implies
that $H$ reaches $3$, but we continue the search by visiting the
next child $I$ to find other possible paths to $3$. Finally, the
algorithm returns the list $(3,L,H,I,1)$.  

While we are finding the list of the nodes to be merged, we also
keep track of the largest SCC, since the SCC with the largest
size is chosen as the new representative.  In our case, SSC $3$,
with $size(3)=5$ is the largest, so all the other nodes are added
under SCC $3$.  This is shown with thick/gray dashed edges in
Figure~\ref{fig:SCC}.  Also note that after this operation, the
nodes $(L,H,I,1)$ are no longer current, and will be added to the
expired SCC nodes $V^e$ (although, as an optimization only the
non-single-node component, namely SCC $1$, is added to $V^e$, and
$L,H,I$ revert to being simple input nodes).  Further note that
we create a new component node only when all the merged
components are input nodes. For the final step, \alg scans the
nodes of the list and updates the DAG edges $E^d$.  The
complexity of this operation is $O(m')$ where $m'$ is the total
number of edges among the nodes to be merged.

\newcommand{\onelabelnode}[2]{$\stackrel{{\stackrel{#2}{~}}}{#1}$}

\begin{figure}[!h]
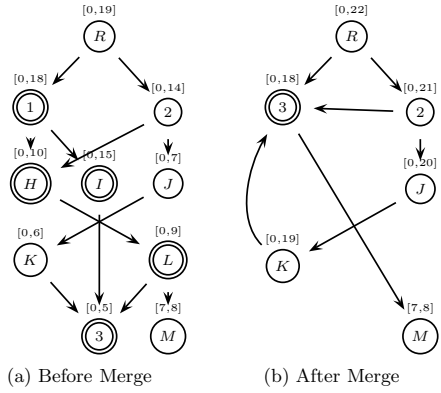

	\vspace{-0.1in}
	\scriptsize
\scalebox{0.8}{%
\centerline{
 \subfigure[Before Merge]{\label{fig:start}
 \psset{xunit=0.45in, yunit=0.5in}
\pspicture[](-1,-0.5)(2.5,4.5)
 \psset{arrows=->,arrowscale=1.5,nodesep=6pt}
 \rput(1,4){\onelabelnode{\circlenode{R}{R}}{[0,19]}}
 \rput(0,3){\onelabelnode{\circlenode[doubleline=true]{A}{1}}{[0,18]}}
  \rput(2,3){\onelabelnode{\circlenode{B}{2}}{[0,14]}}
  \rput(0,2){\onelabelnode{\circlenode[doubleline=true]{C}{H}}{[0,10]}}
  \rput(1,2){\onelabelnode{\circlenode[doubleline=true]{D}{I}}{[0,15]}}
  \rput(2,2){\onelabelnode{\circlenode{E}{J}}{[0,7]}}
  \rput(0,1){\onelabelnode{\circlenode{F}{K}}{[0,6]}}
  \rput(2,1){\onelabelnode{\circlenode[doubleline=true]{G}{L}}{[0,9]}}
  \rput(1,0){\onelabelnode{\circlenode[doubleline=true]{H}{3}}{[0,5]}}
  \rput(2,0){\onelabelnode{\circlenode{I}{M}}{[7,8]}}
  \ncline{R}{A}
  \ncline{R}{B}
  \ncline{A}{C}
  \ncline{A}{D}
  \ncline{B}{E}
  \ncline{C}{G}
  \ncline{E}{F}
  \ncline{G}{H}
  \ncline{G}{I}
  \ncline{B}{C}
  \ncline{D}{H}
  \ncline{F}{H}
\endpspicture
}
 \subfigure[After Merge]{\label{fig:merge}
 \psset{xunit=0.45in, yunit=0.5in}
\pspicture[](-1,-0.5)(2.5,4.5)
 \psset{arrows=->,arrowscale=1.5,nodesep=6pt}
 \rput(1,4){\onelabelnode{\circlenode{R}{R}}{[0,22]}}
 \rput(0,3){\onelabelnode{\circlenode[doubleline=true]{A}{3}}{[0,18]}}
  \rput(2,3){\onelabelnode{\circlenode{B}{2}}{[0,21]}}
  \rput(2,2){\onelabelnode{\circlenode{E}{J}}{[0,20]}}
  \rput(0,1){\onelabelnode{\circlenode{F}{K}}{[0,19]}}
  \rput(2,0){\onelabelnode{\circlenode{I}{M}}{[7,8]}}
  \ncline{R}{A}
  \ncline{R}{B}
  \ncline{B}{E}
  \nccurve[angleA=135,angleB=-120]{F}{A}
  \ncline{E}{F}
  \ncline{B}{A}
  \ncline{A}{I}
\endpspicture
}
}}
\vspace{-0.1in}
\caption{\small Merge Operation on the Index}
\vspace{-0.1in}
\label{fig:indexonmerge}
\end{figure}

\subsubsection{Label Maintenance}
We now discuss how the interval labels are assigned and propagated due to
possible component merge due to the edge insertion.

\smallskip
\noindent{\bf Insertion of DAG Edge:} If the interval of $s$,
$L_s$, already subsumes the interval of $t$, $L_t$, no labels
should be updated.  In fact, this edge insertion eliminates some
false positives which in turn improves the quality of the index.
Otherwise, $L_s = [b_s,e_s]$ is enlarged to cover $L_t$
with $b_s=\min (b_s,b_t)$ and $e_s=\max (e_s,e_t + 1)$.  The
enlargement is propagated up recursively within the DAG to
ensure that each parent contains the intervals of its
children.

\smallskip
\noindent{\bf Merging of Components:} After updating the \alg
graph, we need to assign a label to the representative node for
the merged component.

\begin{remark}
	The SCC
	maintenance algorithm returns a list $l$ of components to be
	merged, as a result of the
	insertion of edge $(s,t)$. The the first node of $l$ is $s$
	whereas the last node is $t$.  Every node in $l$ is reachable
	from $t$ by definition, therefore $L_t$ already subsumes the
	intervals of all the other nodes of $l$.
	\label{rem:mergelabel} 
\end{remark}

We assign the label of the last node of $l$ to the new
representative node due to remark~\ref{rem:mergelabel}.  For
example after the insertion of $(N,B)$, 
the list of component nodes to be merged is $l = (3,L,H,I,1)$, and 
$3$ is chosen as the representative
node. In
Figure~\ref{fig:merge}, we show the labels after the merge
operation.  Node $3$, which is the new representative, 
copies the interval of $1$ which is
$[0,18]$. 

After the merge operation, some parents of the representative
node $c$ might no longer subsume $L_c^i$. In our example, $K$ can
reach $3$, but the previous value of $L_K$, $[0,6]$ in
Figure~\ref{fig:start}, does not contain the new value of $L_3$,
thus its label $L_K$ is enlarged to $[0,19]$ in
Figure~\ref{fig:merge}. Thereafter, $L_J$ is enlarged to $[0,20]$
to cover $L_K$, followed by enlarging $L_2$ to $[0,21]$ and $L_R$
to $[0,22]$.  Essentially, we first recursively update the
ancestors whose start value is larger than $b_c^i$.  Next, we
update the end values of the ancestors, which is different
because end values have to be set to a larger value than its
children. Thus, if there exists more than two paths to a node
$p$, $e_p^i$ and the end values of the ancestors of $p$ might
need to be updated more than once. For instance, if we had
enlarged $L_2$ to $[0,19]$ before enlarging $L_K$, we would have
to enlarge it and its ancestors labels once more because
enlarging $L_K$ also requires the enlargement of $L_2$.  To avoid
this before we update $e_p^i$, we should have updated the end
values of all the children of $p$. We implement this by using a
priority queue which is based on the previous values of the end
values of the nodes. Since a node has a larger end value than its
descendants, it is guaranteed that before we update $p$ we will
have had updated the children.


\textbf{Computational Complexity: } In the worst case, edge insertion
is composed of a reachability query and update of the labels
of the ancestors of the source node. Therefore the computational
complexity is $O(km')$ where $m'$ is the number of edges
in the existing DAG, $|E^d|$. However the propagation 
usually stops after updating a small number of ancestors. If
the source already contains the target node and the new edge
does not merge SCC nodes, the update time is constant.

\begin{figure}[!ht]
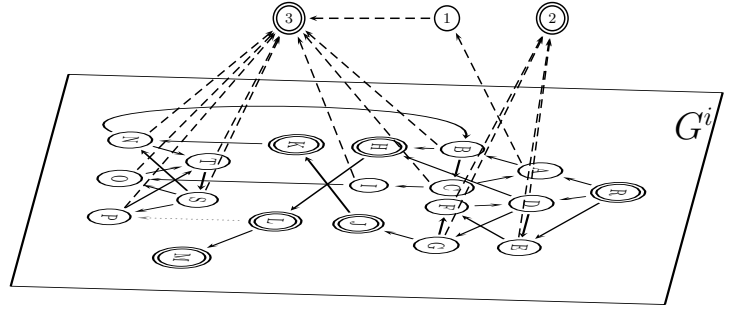

\small
\hspace{0.1in}
\scalebox{0.7}{
\pspicture[](-3.5,-4.5)(3.5,3.5)
 \psset{arrows=->,arrowscale=1,nodesep=2pt}
\rput(8,1.5){\huge $G^i$}
\vshearbox{1.0}{0.90}{
\rotatebox{90}{
 \pspolygon[](-4,-3)(3,-3)(3,5)(-4,5)
 \psset{xunit=0.25in, yunit=0.4in}
 \rput(-1,4){\circlenode[doubleline=true]{R}{R}}
 \rput(-2,3){\circlenode{A}{A}}
 \rput(-3,2){\circlenode{B}{B}}
 \rput(-1,2){\circlenode{C}{C}}
 \rput(-0.3,3){\circlenode{D}{D}}
 \rput(2,3){\circlenode{E}{E}}
 \rput(0,2){\circlenode{F}{F}}
 \rput(2,2){\circlenode{G}{G}}
  \rput(-3,1){\circlenode[doubleline=true]{H}{H}}
  \rput(-1,1){\circlenode{I}{I}}
  \rput(1,1){\circlenode[doubleline=true]{J}{J}}
  \rput(-3,0){\circlenode[doubleline=true]{K}{K}}
  \rput(1,0){\circlenode[doubleline=true]{L}{L}}
 \rput(-3,-2){\circlenode{N}{N}}
 \rput(-1,-2){\circlenode{O}{O}}
 \rput(1,-2){\circlenode{P}{P}}
 \rput(-2,-1){\circlenode{T}{T}}
 \rput(0,-1){\circlenode{S}{S}}
  \rput(3,-1){\circlenode[doubleline=true]{M}{M}}
  \ncline{R}{A}
  \ncline{R}{D}
  \ncline{R}{E}
  \ncline{A}{B}
  \ncline{B}{C}
  \ncline{C}{A}
  \ncline{D}{E}
  \ncline{F}{D}
  \ncline{D}{G}
  \ncline{E}{F}
  \ncline{G}{F}
  \ncline{B}{H}
  \ncline{C}{I}
  \ncline{G}{J}
  \ncline{D}{H}
  \ncline{H}{L}
  \ncline{J}{K}
  \ncline{L}{M}
  \ncline{K}{N} 
  \ncline{I}{O}
  \ncline[linestyle=dotted,dotsep=1pt,linecolor=gray]{L}{P}
  \ncline{N}{T}
  \ncline{O}{T}
  \ncline{P}{T}
  \ncline{T}{S}
  \ncline{S}{N}
  \ncline{S}{O}
  \ncline{S}{P}
  \nccurve[angleA=-135,angleB=180]{N}{B}
}
}
 \rput(0,3.5){\circlenode[doubleline=true]{3}{3}}

 \rput(3,3.5){\circlenode{1}{1}}
 \ncline[linestyle=dashed]{A}{1}
 \ncline[linestyle=dashed]{B}{3}
 \ncline[linestyle=dashed]{C}{3}

 \rput(5,3.5){\circlenode[doubleline=true]{2}{2}}
 \ncline[linestyle=dashed]{D}{2}
 \ncline[linestyle=dashed]{E}{2}
 \ncline[linestyle=dashed]{F}{2}
 \ncline[linestyle=dashed]{G}{2}

 \ncline[linestyle=dashed]{S}{3}
 \ncline[linestyle=dashed]{T}{3}
 \ncline[linestyle=dashed]{O}{3}
 \ncline[linestyle=dashed]{N}{3}
 \ncline[linestyle=dashed]{P}{3}

 \ncline[linestyle=dashed]{1}{3}
 \ncline[linestyle=dashed]{I}{3}
\endpspicture
}
\vspace{-0.75in}

\caption{\small Deletion of (gray dotted) edge $(L,P)$ from $G^i$.
First, node $L$ becomes a component by itself, and we remove the
containment link $(L,3)$. Then $H$ does the same. The call from
$B$ finds a path to the target node $P$ via $C,I,O,T,S$ therefore
these nodes remain under $3$ along with $P$. Note that we do not
touch $A$ and $N$ since they will continue to remain under $3$.
Also the containment edges $(B,1)$ and $(C,1)$ are removed.
Containment edges $(B,3)$ and $(C,3)$ are added when we lookup
for their component, due to path compression within the
union-find structure. However $(A,1)$ is remains unchanged. To
reduce clutter, the DAG edges $E^d$ are not shown.}

\label{fig:Deletion}
\end{figure}
\vspace{-0.1in}
\subsection{Edge Deletion}
\label{sec:edgedeletion}
\subsubsection{SCC Maintenance}
\label{sec:extractcomponents}
Nodes of $G^i$ have to be examined to detect the consequences of
edge deletions on the \alg graph $G$.  Deletion of $e=(u,v)$ may
cause a split of an existing component only if $u$ and $v$ are in
the same component. If they are in different components $s$ and
$t$, we just check whether the removal of $e$ also removes the
$(s,t)$ edge at the DAG level. If that is the case we remove the
$(s,t)$ edge and update the labels.

If both $u$ and $v$ are members of the same component $s$, one
naive way to find the emerging components is to  perform Tarjan's
Algorithm on the nodes within SCC $s$.  This algorithm would work
well if the sizes of the strongly connected components are
relatively small.  The complexity of the method is $O(m')$ where
$m'$ is the number of edges inside the component $s$. However in
real-world graphs, it is not uncommon to observe giant strongly
connected components with size $O(n)$. Furthermore, the nodes
inside this component are usually highly connected and removal of
edges are less likely to split the component. Even if it breaks
up, the expectation is that there will still be a relatively
large strongly connected component remaining. Therefore our goal
is to devise an algorithm which can extract the new components
without traversing all nodes, especially the ones that are going
to stay within the large component. 

\begin{remark} When we delete the edge $e=(u,v)$ from the
	component $s$, we know that there still exists at least one
	path from $v$ to $u$.  If we can still reach from $u$ to $v$,
	it means that they are still in the same component and the
	other members of $s$ are also remain unchanged.
	\label{rem:unchanged} \end{remark}

Our algorithm is based on the fact that node $v$ reaches every
other node in $s$. We want to find the new components without
visiting the nodes that are going to stay in the same component
as $v$ after the split. We first start a traversal from the node
$u$. From remark~\ref{rem:unchanged}, if we find a path to $v$,
the algorithm terminates without changing the component $s$.
However if there is no such path, there must be at least one new
component which contains $u$. We create new SCC nodes for these
new components. To find the the new components that are reachable
from $s$, we use Tarjan's algorithm with the following
modifications:  i) we only traverse the nodes in $s$, ii) if we
find that the node reaches the target node $v$, we can
immediately return without visiting other children, and iii) if a
new component is found, we push all of the parents of the nodes
of this new component to a queue.  This same approach is applied
for each node in the queue.  The reason we add those parent nodes
to the queue is that they may become a part of another component
which does not reach $v$. However we do not add the parents of a
node $w$ if $w \rightsquigarrow v$, because it implies that all
the ancestors of $w$ can reach $v$ via $w$ which makes them
remain in the same component with $v$, which is $s$ (note that $v
\rightsquigarrow w$ using a similar argument as
remark~\ref{rem:unchanged}).

In a nutshell, components are extracted in a bottom-up manner for
all nodes $w$ unless we are sure that $w$ can reach $v$.  The
main benefit of this algorithm as opposed to the naive approach
is in the pruning. It may find out the components without
traversing all the members of $s$. Finally, we create a new node
in $V^c$ for each component if it is not a single node component,
and we add the new node to the result list.

\begin{figure}[!h]
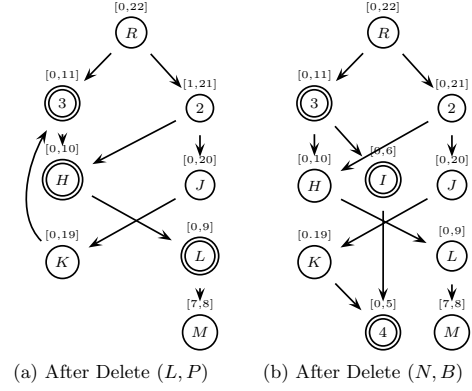

	\scriptsize
	\scalebox{0.8}{
	\centerline{
 \subfigure[After Delete ($L,P$)]{\label{fig:splitLP}
 \psset{xunit=0.45in, yunit=0.5in}
\pspicture[](-1,-0.5)(2.5,4.5)
 \psset{arrows=->,arrowscale=1.5,nodesep=6pt}
 \rput(1,4){\onelabelnode{\circlenode{R}{R}}{[0,22]}}
 \rput(0,3){\onelabelnode{\circlenode[doubleline=true]{A}{3}}{[0,11]}}
  \rput(2,3){\onelabelnode{\circlenode{B}{2}}{[1,21]}}
  \rput(0,2){\onelabelnode{\circlenode[doubleline=true]{C}{H}}{[0,10]}}
  \rput(2,2){\onelabelnode{\circlenode{E}{J}}{[0,20]}}
  \rput(0,1){\onelabelnode{\circlenode{F}{K}}{[0,19]}}
  \rput(2,1){\onelabelnode{\circlenode[doubleline=true]{G}{L}}{[0,9]}}
  \rput(2,0){\onelabelnode{\circlenode{I}{M}}{[7,8]}}
  \ncline{R}{A}
  \ncline{R}{B}
  \ncline{A}{C}
  \ncline{B}{E}
  \ncline{C}{G}
  \ncline{E}{F}
  \ncline{G}{I}
  \ncline{B}{C}
  \nccurve[angleA=135,angleB=-120]{F}{A}
\endpspicture
}
 \subfigure[After Delete ($N,B$)]{\label{fig:splitNB}
 \psset{xunit=0.45in, yunit=0.5in}
\pspicture[](-1,-0.5)(2.5,4.5)
 \psset{arrows=->,arrowscale=1.5,nodesep=6pt}
 \rput(1,4){\onelabelnode{\circlenode{R}{R}}{[0,22]}}
 \rput(0,3){\onelabelnode{\circlenode[doubleline=true]{A}{3}}{[0,11]}}
  \rput(2,3){\onelabelnode{\circlenode{B}{2}}{[0,21]}}
  \rput(0,2){\onelabelnode{\circlenode{C}{H}}{[0,10]}}
  \rput(1,2){\onelabelnode{\circlenode[doubleline=true]{D}{I}}{[0,6]}}
  \rput(2,2){\onelabelnode{\circlenode{E}{J}}{[0,20]}}
  \rput(0,1){\onelabelnode{\circlenode{F}{K}}{[0.19]}}
  \rput(2,1){\onelabelnode{\circlenode{G}{L}}{[0,9]}}
  \rput(1,0){\onelabelnode{\circlenode[doubleline=true]{H}{4}}{[0,5]}}
  \rput(2,0){\onelabelnode{\circlenode{I}{M}}{[7,8]}}
  \ncline{R}{A}
  \ncline{R}{B}
  \ncline{A}{C}
  \ncline{A}{D}
  \ncline{B}{E}
  \ncline{C}{G}
  \ncline{E}{F}
  \ncline{G}{I}
  \ncline{B}{C}
  \ncline{F}{H}
  \ncline{D}{H}
\endpspicture
}
}}
\vspace{-0.1in}
\caption{\small Split Operations on the Index}
\vspace{-0.1in}
\label{fig:indexonsplit}
\end{figure}

As an illustration of the algorithm, we delete the edge $(L,P)$
in Figure~\ref{fig:Deletion} from the final graph we obtained in
Figure~\ref{fig:SCC}. Since they are in the same component,
we call the extract component algorithm from
node $L$. We skip the child $M$ as it is not a member of $3$.
Since $L$ has no other child, the function returns after
putting its parent $H$ into the queue. Similarly the call from $H$
returns after inserting its parent $B$ into the queue. When we run
the recursive method from $B$, the function will recurse into the
nodes $C,I,O,T,S$ and $P$. As soon as it finds $P$, it will
backtrack to $B$ marking each of the visited nodes.
If a node is marked, we know for sure that
it is in the same component as the target node. Furthermore,
we do not put their parents into the queue. For this reason,
the algorithm does not touch the nodes $A$ and $N$ in our example.
That is also why $A$ still has a containment edge to node $1$.
Finally, 
the resulting list contains the new components $L$ and $H$. 
The DAG after deletion of $(L,P)$ is shown in
Figure~\ref{fig:splitLP}, whereas the complete \alg graph is shown
in Figure~\ref{fig:Deletion}

If we delete $(N,B)$ as a second example, we first extract the
components from $N$, which finds a new component comprised of
$(N,T,O,S,P)$.  Since it is not a single node component we create
a new SCC node $4$ to represent them. We add parents of these
nodes which are member of SCC $3$ into the queue as potential new
components. These are $I$ and $B$. When we recursively call the
extract component method we find that $I$ is a single node
component, as it has only one child which is already visited.
Finally, the routine stops at $B$ since it is the target node.
$B$ and its ancestors (which are $A$ and $C$) are left in SCC $3$
without processing.  The algorithm returns the list $(4,I,3)$.
See Figure~\ref{fig:splitNB} for the final DAG after deletion of
$(N,B)$.

\subsubsection{Label Maintenance}

\noindent{\bf Removal of a DAG Edge:} If an edge $(s,t)$ is
removed from the DAG, we do not have to update the interval
labels. This is because the interval of $s$ will still contain
the interval of $t$, which only introduces some false positives
without invalidating the index. However, to avoid some of those
false positives the label of $s$ can be shrunk if the interval of
$t$ is at the beginning or at the end of $L_s$.  Furthermore, the
same change should be applied to parents of $s$ recursively as
long as it is possible to shrink the interval of the parent node.
We use the simple strategy in this paper, namely, we do not
update the labels when an edge is removed from $E^d$.

\smallskip \noindent{\bf Splitting of Components:} Upon the
deletion of the edge $(u,v)$ from the component $s$, if the
component breaks up, the component extraction algorithm (in
Section~\ref{sec:extractcomponents}), returns a list (called
$clist$) of new components.  These nodes constitute a directed
acyclic graph which has a single root $s$, and a single leaf $t$
(i.e., the component that contains $u$).  We perform random
traversals (as we do for initial assignments) that visit only the
nodes of $clist$.  The algorithm starts assigning intervals from
$b_s^i$ and uses the size of new components in computing end
values.  Consequently, since all paths in $clist$ lead to node
$t$, $L_t^i$ gets assigned $[b_s^i,b_s^i+size(t)]$ unless it has
an outgoing edge to other components which have a larger end
value.  Therefore,  we keep a counter $ctr$ which provides the
value of the post-order value of a node. We also maintain the
largest end value $end$ of the children.  The counter values are
incremented by size of the component node. The final interval for
a node $u$ ends at the larger of the values of $ctr$ and $end+1$.
There is a slight possibility that the interval of a node becomes
larger than the former interval of $s$. In that case the label
has to be propagated up.

In Figure~\ref{fig:indexonsplit}, we show two edge deletions that
cause split operations. When $(L,P)$ is deleted, $H$ and $L$ are
separated from $3$ (see also Figure~\ref{fig:Deletion}).  A
post-order traversal would assign $[0,1]$, $[0,2]$ and $[0,11]$
to $L$, $H$ and $3$, respectively. However due to edge $(L,M)$ we
enlarge $L_L$ to $[0,9]$, $L_M$ to $[0,10]$ and finally $L_3$ to
$[0,11]$.  Since none of the new labels exceed the old interval
of $3$, we do not need to propagate them further. Next we delete
the edge $(N,B)$, which results in the separation of $4$ and $I$
from $3$. Note that after the split, $size(3)$ is 3 and $size(4)$
is 5.  A post-order traversal would assign $[0,5]$, $[0,6]$ and
$[0,9]$ to the nodes $4$, $I$ and $3$ respectively. However due
to edge $(3,H)$ we keep $[0,11]$ for node $3$.

\textbf{Computational Complexity: } The cost of deletion is
constant if source and target nodes are in different components.
Otherwise the cost is $O(m'-m^{''}+n^{''})$ where $m'$ is the
edge size of the component before split and $m^{''}$ and $n^{''}$
are the edge and node sizes of the component where the source
node resides after the split. In fact, this is efficient
especially when components are tightly connected internally,
which is usually the case in real graphs, such as web graphs and
social networks.

\subsection{Node Insertion and Deletion}
\label{sec:nodeinsertion}

When a node $u$ is inserted with incoming edge list $l_i$ and
outgoing edge list $l_o$, we first add $u$ to $G^i$ with its
outgoing edges. Since we haven't processed any of its incoming
edges yet, $u$ cannot be a member of a larger strongly connected
component. Therefore $u$ is a SCC node with component size $1$.
$[b_u,e_u]$ is assigned to $L_u$ where $\displaystyle b_u =
\min_{w\in l_o}(b_{S(w)})$ and $\displaystyle e_u = \max_{w\in
l_o}(e_{S(w)})+1$, where $S(w)$ is the component id for node $w
\in l_o$. If it has no outgoing edges $b_u$ is set to the largest
existing end value and $e_u$ is set to $b_u+1$. After that the
incoming edges $l_i$ are inserted one by one via the edge
insertion algorithm.  Thus the computational complexity of node
insertion is $O(k|l_o| + |l_i|C_{ei})$ where $C_{ei}$ is the edge
insertion time.

The deletion of nodes can also be defined in terms of edge
deletions.  When we are deleting node $u$, we first delete all of
its outgoing edges one by one via the edge deletion algorithm in
Section~\ref{sec:edgedeletion}. Once all its outgoing edges are
deleted, it becomes an SCC node with size 1. All the incoming
edges of $u$ now become inter-component edges whose removal does
not change the labels of other SCC nodes. Thus we do not invoke
the edge deletion method for these.  Lastly, we remove the node
$u$ from $G^i$. Thus, the cost of node deletion is
$O(|l_o|C_{ed})$ where $|l_o|$ is the outdegree of the node and
$C_{ed}$ is the cost of single edge deletion.

\subsection{Batch Update Operations}

DAGGER is especially useful for graphs that undergoes several reachability
queries between each atomic updates. The reason is that DAGGER is specifically 
designed for fine-grained update operations and thus the index can always reflect
the underlying graph consistently. However in some scenarios graphs can receive
update operations in batches. For example, an online social network service may want
to report to its database after accumulating updates for certain period of timeframes 
instead of having round trips to the database at every single graph update.
The obvious solution to handle a batch of $K$ updates is subsequently processing
them. This may become quite forbidding for large values of $K$ as the runtime of 
an update operation can be linear on the graph size in the worst case. In this section, 
we discuss on how we can handle a batch of $K$ updates better than independently
processing $K$ single updates.

To begin with it is worth noting that reconstructing the index from scratch is the best solution
if the size of the update graph is in the order of the size of original graph. Because DAGGER 
index construction is a lightweight operation and could definitely be faster and healthier 
than incorparating the atomic graph updates into the index one by one. For relatively smaller batches,
we can apply the following pruning and reordering techniques. For the following discussion, 
we assume update set $U$ is composed of $K$ update operation, each of which we categorize in four
classes. We notate an update on edge $e$ as $e^+$(or $e^-$) if it is an insertion (or deletion).
Similarly we use the subscripts $e_\curvearrowright$ and $e_\circlearrowleft$ to indicate if the 
edge is an inter-component or intra-component edge respectively. Therefore we have four disjoint sets of
 edges composing $ U = \left\{ e_1, \dots, e_K | e_i \in 
U^+_\circlearrowleft \cup U^-_\curvearrowright \cup U^-_\circlearrowleft \cup U^+_\curvearrowright \right\} $.

\begin{list}{\labelitemi}{\leftmargin=1em}
 \item {\textbf Pruning:} Find pairs of complementing update operations (e.g., insertion and deletion of 
a specific edge $e$) and remove them from the update list.
 \item {\textbf Preprocessing:} The condensation of the update graph can be found as in DAGGER construction.
That would reduce the number of insertion operations in the update list.
 \item {\textbf Batching:} We can take advantage of batching if we group edges by type and process them 
in the following order.
 \begin{list}{\labelitemi}{\leftmargin=1em}
  \item {$U^+_\circlearrowleft$:} These edges can be added into graph with no changes to the index,
as they do not change the reachability of the graph.
  \item {$U^-_\curvearrowright$:} We can also skip the inter-component edge removals just by updating
the graph and not modifying the index as we allow false positives.
  \item {$U^+_\curvearrowright$:} The real gain of batching happens in this type of updates. Normally
insertion of an inter-component edge may cause enlargement of a label which should be propagated up
in DAG $G^d$. Therefore $k$ such updates have complexity $O(k|G^d|)$. However this can be reduced to
$O(k+|G^d|)$ if all $k$ edges are added first into $G^i$ at once and propagating labels up in the DAG
in one pass.
  \item {$U^-_\circlearrowleft$:} A similar approach does not work for these kind of updates
since remark~\ref{rem:unchanged} does not hold when we delete multiple edges from a SCC at a time.
 \end{list}
\end{list}

\section{Experiments}

In this section, we evaluate the effectiveness of DAGGER on various
real and synthetic datasets. We compare DAGGER with 
the baseline Depth-First
Search (DFS).
The other methods are not included in our comparison 
because of the following issues with them:
i) Optimal-Tree Cover~\cite{agrawal}: Although it mentions
how to update the index for some operations, it does not support
all operations. Furthermore, it assumes the graph is always acyclic.
ii) Incremental-2HOP~\cite{bramandiaJ}: The existing
implementation do not support all update operations. It
supports edge insertions and node deletions.
iii) \cite{krommidas2008} provides an experimental analysis 
and implementations of the dynamic transitive closure
\cite{roditty,king-jcss02,king,henzinger,demetrescu2000}
approaches. However,
none of them are scalable as they require quadratic space.  

In these experiments, we attempt to index very large dynamic
graphs on a system with a quad-core Intel i5-2520M 2.50Ghz
processor, with
4GB memory. To the best of our knowledge, the largest 
dynamic graphs previously used for reachability queries had 
400K nodes~\cite{bramandiaJ}.
Furthermore, that study only applied node deletions.
In that sense, our 
study is unique in that it scales to million node graphs, and
includes a comprehensive evaluation over an intermixed sequence
of update operations, i.e., our evaluation includes all four
update operations -- edge/node insertion/deletion.

\subsection{Experimental Setup}
To effectively and realistically
evaluate the cost of index maintenance, we intersperse
reachability queries with the graph update operations.
We then compare \alg with DFS to find out
under which conditions \alg's fast querying 
amortizes the maintenance cost of the index.
Note that DFS has no update cost, but pays a penalty in terms of  longer query 
times. Thus, in a scenario that receives queries
very rarely (e.g., 1 query per hundreds of updates),
it is obvious that maintaining an index would not pay
off. Thus, in our experiments we vary the ratio of queries
to update operations, and measure the total time which
includes the query and update times for both DFS and \alg.
For each dataset, we report the average time for the update
operations and queries, as well as the
total time taken for $q$ queries per update (QpU), 
where $q$ ranges from one to eight.  As we shall see, \alg
typically outperforms DFS after just 2 queries per update, with
the gap increasing as this ratio goes up. This 
makes it an effective index for dynamic real-world graph querying.

\begin{table}
\centering
\scalebox{0.8}{
 \begin{tabular}{ll|r|r|r|r}
\multicolumn{2}{c|}{Graph} & Node & Edge & DAG & Largest \\
\multicolumn{2}{c|}{Dataset} & Size & Size & Size & SCC\\
\hline
\multirow{2}{*}{FrenchWiki} &Initial & 999,447 & 3,452,667 & 899,343  &  97,465\\
& Final & 999,499 & 3,452,953 & 899,360 & 97,502 \\
\hline
\multirow{2}{*}{PatentCitation}& Initial & 605,617 & 1,000,002 & 605,617  &  1\\
& Final & 705,617 & 1,205,190 & 705,617 & 1 \\
\hline
\multirow{2}{*}{ER1M}& Initial & 1,000,000 &1,500,123 & 659,892 & 340,109\\
& Final & 1,000,170 & 1,500,631 & 659,647 & 340,505 \\
\hline
\multirow{2}{*}{BA1M}& Initial & 1,000,000 &2,000,823 &  478,219 & 521,782 \\
& Final & 1,000,182 & 2,001,313 & 478,240 & 521,930
\end{tabular}
}
\vspace{-0.1in}
\caption{\small Properties of dynamic datasets: Node size refers
to $|V^i|$, edge size to $|E^i|$, and DAG to $|V^d|$; Largest SCC
refers the component $size$, i.e., number of nodes in the largest SCC.}
\label{tab:dynamicdata}
\end{table}

\subsection{Datasets}

As opposed to the static setting, evaluation of dynamic indexing
requires a valid sequence of update operations, as well as the
queries, along with an initial graph. We used both real and
synthetic graph datasets in our experiments, as shown in
Table~\ref{tab:dynamicdata}.  The row marked ``Initial'' shows
the properties of the input graphs, whereas the row marked
``Final'' shows the same properties after the completion of all
the update operations.

\smallskip\noindent
\textbf{Real Graph Evolution: } For these datasets we were
able to compile the complete evolution of a graph at the edge
level. We have two such datasets:
\begin{list}{\labelitemi}{\leftmargin=1em}
 \item {FrenchWiki}: Wikimedia Foundation dumps the snapshots
	 of Wikipedia in certain intervals (see
	 \url{dumps.wikimedia.org}).
These dumps contain all the textual content with full
revision/edit history. We discarded the textual content
and recovered the evolution of the complete French language Wikipedia graph from
its birth by comparing the consecutive versions of
each page. If a new wiki-link is added in a version
of a wiki-page, we consider it as an insertion of
an edge with the timestamp of that version of the wiki-page. 
Similarly, if an existing wiki-link disappears in a version
of a wiki-page, we consider it as an edge deletion.
We took a snapshot of the graph and indexed it
when it had 1 million nodes, and we then applied the next 1000
update operations to the indexed graph.
\item {PatentCitation}: This graph 
includes all citations 
within patents granted in the US between 1975 and 1999 (see
\url{snap.stanford.edu/data/cit-Patents.html}). Since
the timestamps of the patents are also available, so we can simulate
the growth of the data. Note that the only update in this data is 
node additions with a set of outgoing edges. Therefore it is 
always an acyclic graph. We indexed a snapshot
of the graph when it had around 600K nodes, and we then 
applied the following 100,000 node additions
to the initial graph as update operations.
\end{list}

\smallskip\noindent
\textbf{Synthetic Graph Evolution: } For these datasets, we first
generated three different random graphs. 
We then generated a synthetic
update sequence of 1,000 operations. Our random graphs are:
\begin{list}{\labelitemi}{\leftmargin=1em}
\item {ER1M}: We generated a directed graph of 1 million nodes, and
	1.5 million edges using the Erdos-Renyi (ER)~\cite{ER} random
	graph model.  Each edge is selected by choosing a source and
	a target node, both uniformly at random, and directing the
	edge from the source to the target.
\item {BA1M}: The Barabasi-Albert (BA) preferential
	attachment model~\cite{Barabasi1999} is a generative model
	which retains some real-world graph properties such as
	power-law degree distributions. The average degree $d$ is a
	parameter of the model. Starting from $2d$ initial nodes, at
	each time step a new node is added to the graph, with some
	outgoing edges to the existing nodes. The number of outgoing
	edges is chosen randomly in the range [1,2d]. Further, the
	end point for each edge is ``preferentially'' selected with
	probability proportional to the degree of the existing node.
	In other words, when node $w$ is being inserted, it will be
	connected to node $x$ via the edge $(w,x)$ with probability
	$degree(x)/2m$ where $m$ is the current edge size of the
	graph. To obtain possibly cyclic graphs with BA model, we
	reversed the new edges with probability $0.5$. Therefore, the
	generative BA process can create cycles. 
	We generated a graph with 1 million nodes using this directed
	BA model (with $d=2$).
\end{list}

For each of the synthetic graphs, we randomly generated update
sequences using a preferential attachment model.  In the
generation of the update sequence, we first select the operation
type with predefined ratios (for instance, we used 60\% insert
edge, 15\% delete edge, 20\% insert node, 5\% delete node, which
can be considered as representing a insertion focused graph
growth scenario, with more weight on adding edges).
Since we also report the average time for each operation, 
the exact
ratios of different operations does not effect our conclusions.
The sequence of update operations is generated as follows: 
i) Insert Edge: Select source node uniformly at random and target
node via preferential attachment.
ii) Delete Edge: Select an edge from the existing edges uniformly at random.
iii) Insert Node: Randomly determine the indegree and outdegree of the
node, then select the other ends for these edges via preferential 
attachment.
iv) Delete Node: Select a node uniformly at random and delete
it with its incident edges.
 
In our experiments, we measure the average update times (i.e.,
edge insertion time (EI), node insertion time (NI), edge deletion
time (ED) and node deletion time (ND)) during the lifetime of a
dynamic reachability index. We used three versions of \alg --
DG0, DG1 and DG2 correspond to \alg with no intervals ($k=0$), 1
interval ($k=1$), and 2 intervals ($k=2$) per node, respectively.
Note that DG0 maintains only the DAG graph without using any
interval labeling and answers queries by performing a search on
the DAG graph. In contrast, the basic DFS performs the query
directly on the input graph, since it obviously does not maintain
the DAG.

\begin{figure*}[!ht]
\centerline{
\subfigure[FrenchWiki]{
\includegraphics[width=1.35in,height=1.00in]{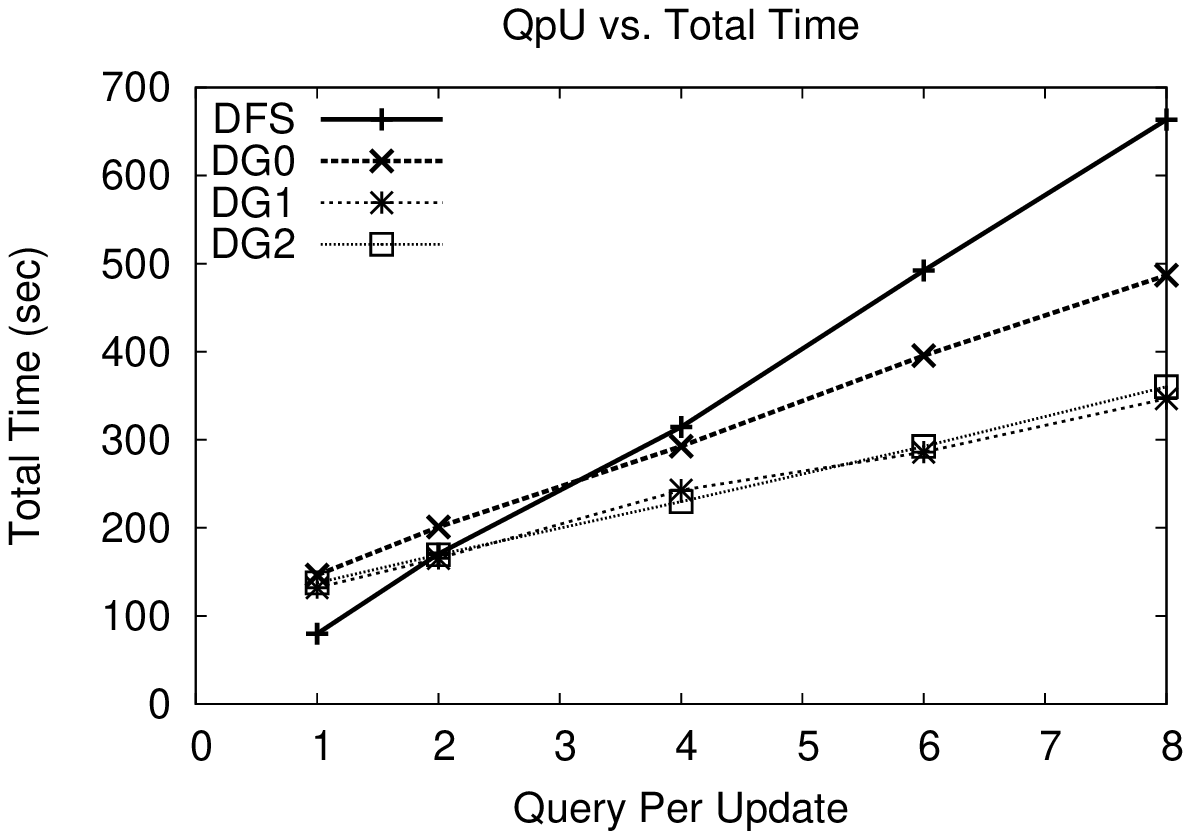}	
}
\subfigure[PatentCitation]{
\includegraphics[width=1.35in,height=1.00in]{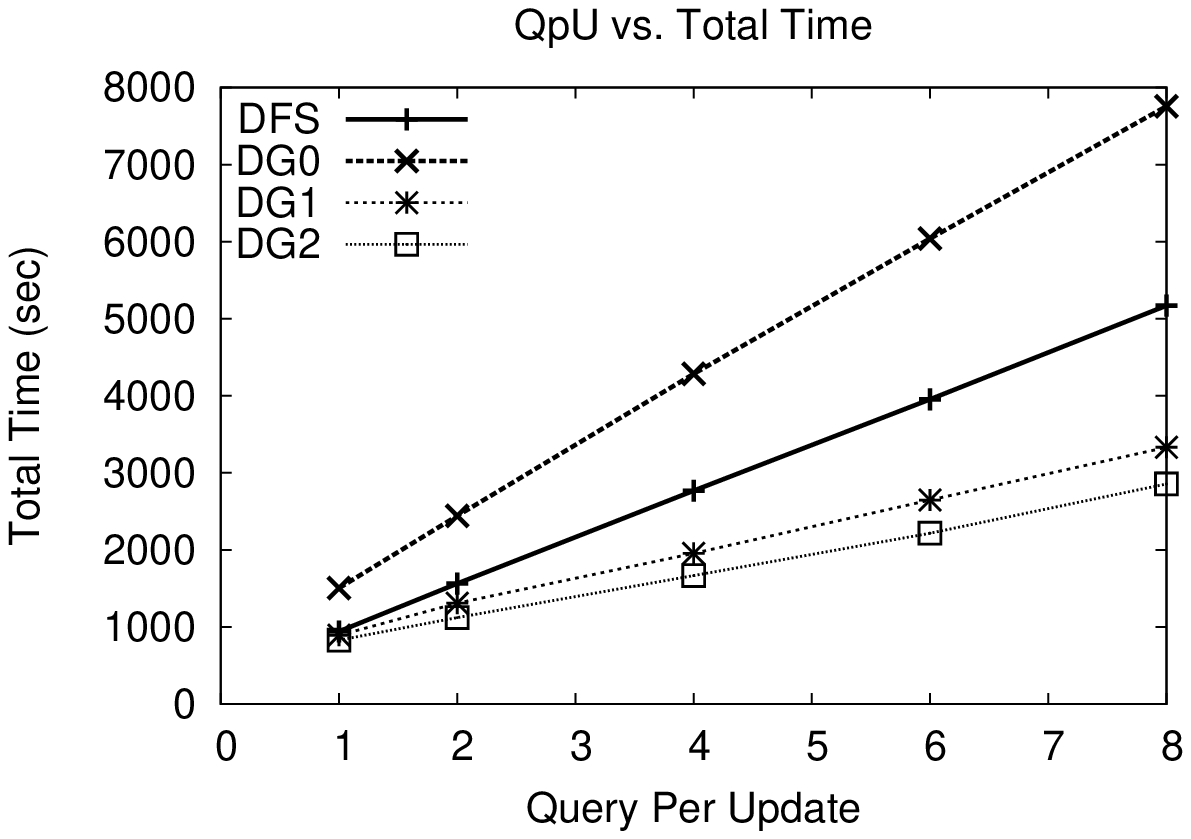}	
}
\subfigure[ER1M]{
\includegraphics[width=1.35in,height=1.00in]{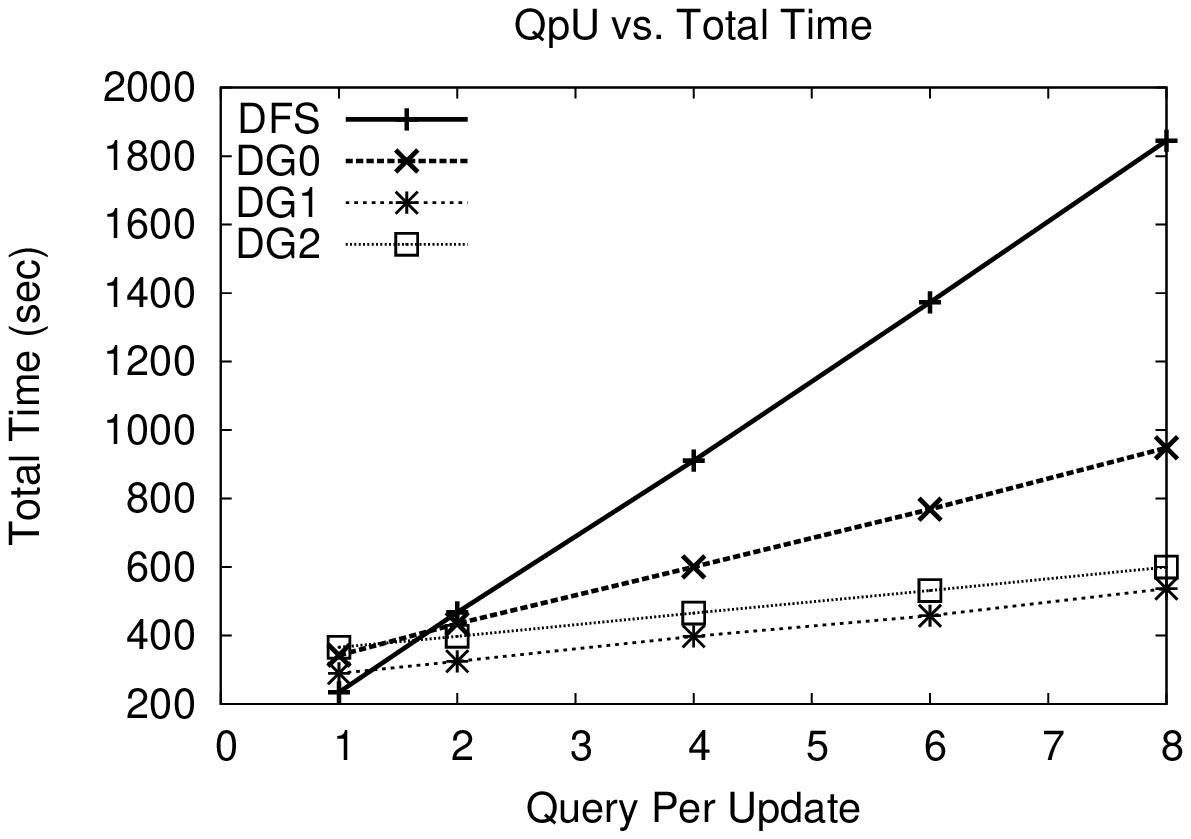}	
}
\subfigure[BA1M]{
\includegraphics[width=1.35in,height=1.00in]{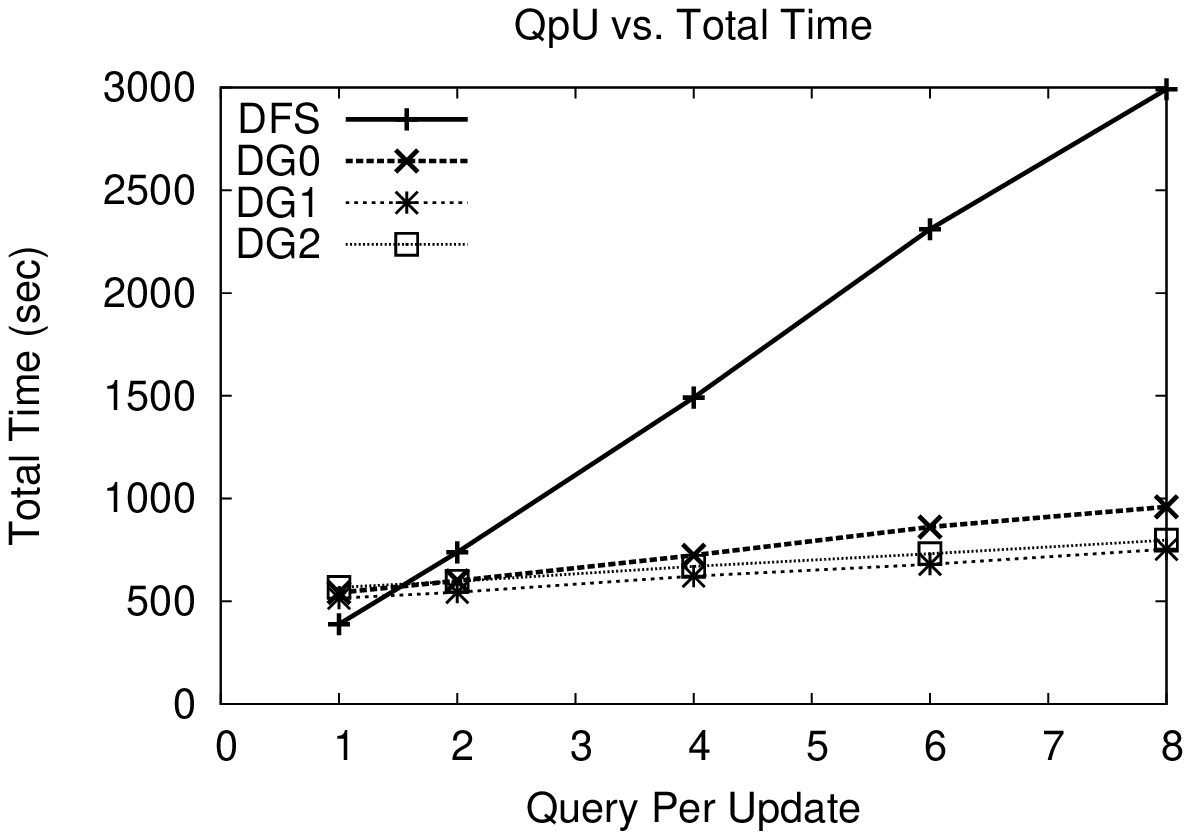}	
}}
\vspace{-0.15in}
\caption{\small Total time comparison on dynamic datasets}
\label{fig:dynamictotal}
\end{figure*}

\begin{table}[!ht]
\centering
\small
\subtable[Real Graphs]{
\begin{tabular}{|l||ccccc||cc|}
\hline
Data & \multicolumn{5}{|c||}{FrenchWiki} & \multicolumn{2}{|c|}{Citation}\\
\hline
Method & Q & EI & ED & NI & ND & Q & NI\\
\hline
DFS & 410 & - & - & - & - & 0.19 & - \\
DG0 & 221 & 60 & 253 & 0.06 & 191 & 0.31 & 0.06 \\
DG1 & 148 & 48 & 267 & 0.02 & 189 & 0.10 & 0.06 \\
DG2 & 128 & 62 & 279 & 0.02 & 192 & 0.08 & 0.06 \\
\hline
\end{tabular}
}
\vspace{0.2in}
\hspace{-0.4in}
\scriptsize
\subtable[Synthetic Graphs]{
\begin{tabular}{|l||ccccc||ccccc|}
\hline
Data & \multicolumn{5}{|c||}{ER1M} & \multicolumn{5}{|c|}{BA1M}\\
\hline
 & Q & EI & ED & NI & ND & Q & EI & ED & NI & ND\\
\hline
DFS & 311 & - & - & - & - & 448 & - & - & - & -  \\
DG0 & 119 & 212 & 775 & 0.01 & 538  & 75 & 172 & 2032 & 0.03 & 2726\\
DG1 & 46 & 203 & 832 & 0.13 & 548  & 39 & 148 & 2138 & 0.13 & 2769\\
DG2 & 44 & 319 & 871 & 0.10 & 562 & 38 & 218 & 2202 & 0.07 & 2804 \\
\hline
\end{tabular}
}
\caption{\small Average Operation Times (in ms) on Real and
Synthetic Data. $Q$ refers to query time, $EI$ and $ED$ to edge
insertion and deletion, and $NI$ and $ND$ to node insertion and
deletion.}
\label{tab:dynamicresults}
\end{table}

\subsection{Results}

In Figure~\ref{fig:dynamictotal}, we plot the total time taken to
perform all the operations (updates and queries) for DFS and 
the \alg variants DG0,
DG1, and DG2. The total time (in sec) is plotted against the
number of reachability queries per update operation. DAGGER has the
advantage of fast querying at the cost of index maintenance,
whereas DFS has no update cost. These plots make it clear that 
DG1 amortizes the maintenance costs as long as there are 
2-4 queries per update operation, which is quite reasonable
for an online system. The average update and query times on 
each of the graph datasets are shown in
Tables~{\ref{tab:dynamicresults}(a)} and {\ref{tab:dynamicresults}(b)}.
We discuss dataset specific results below.

\textbf{FrenchWiki:} In Figure~\ref{fig:dynamictotal}(a),
we see that DG0 performs better DFS in total time starting with 
4 queries per update. 
In other words, the efficient dynamic DAG/SCC maintenance in \alg
is enough to outperform the baseline DFS method.
On the other hand, the interval labeling and maintenance in \alg
offers even more significant benefits.
We can observe that 
\alg with interval labeling is better than DFS after just
2 queries per update. 
However, it is interesting to note that there is no big 
difference in the performance of DG1 and DG2. That is, using two
interval labels does not improve the total time. 
In Table~{\ref{tab:dynamicresults}(a)} we can see that whereas the
average query time is indeed better for DG2, unfortunately it
incurs more overhead for the update operations due to the
extra cost of maintaining one more interval, in particular for
edge insertion/deletion.

\textbf{PatentCitation:} In this dataset the only update
operation is node insertion with outgoing edges. From
Table~{\ref{tab:dynamicresults}(a)} we observe that the cost of node
insertions are very close for all versions of DAGGER due to the
fact that there is no DAG maintenance (e.g., input graph is
always a DAG) and no label propagation. Therefore, the plot in
Figure~\ref{fig:dynamictotal}(b) reflects mainly the query time of
the methods. DG0 has worse query times than DFS, even though it
performs the same search as DFS, which can be attributed to the
overhead of the \alg graph.  However, dynamic interval labeling
provides a significant improvement on querying performance, with
two intervals (DG2) providing better query times than a single
interval label per node (DG1).

\textbf{ER1M: } Table~\ref{tab:dynamicdata} shows that this graph
has a large connected component one-third the size of the input
graph; almost all the remaining nodes are single node components.
Table~{\ref{tab:dynamicresults}(b)} shows the break down in terms of
average operation times.  We observe that DFS querying is three
times slower than even DG0! Further, dynamic interval labeling
helps, but the small query performance gain obtained by utilizing
two intervals per node instead of single interval does not pay
off the label maintenance cost. In particular, edge insertion
with DG2 is significantly slower than DG1, because when the
interval of the huge component is enlarged, this change has to be
propagated up for all the incoming edges of the huge component,
and there can be many such incoming edges.  However, the cost of
edge insertion does not increase when moving from DG0 to DG1,
because interval labels also provide pruning when updating the
DAG. The plot in Figure~\ref{fig:dynamictotal}(c) shows that DG1 is
the best method overall, and all \alg variants are preferable to
DFS, if there are at least 2 queries per update.

\textbf{BA1M: } The results are similar to those
for ER1M (see Table~{\ref{tab:dynamicresults}(b)}).
The main difference is that BA1M has a much larger SCC,
which reduces the size of the DAG significantly. Hence,
there is a greater performance difference between DFS and
\alg methods as seen in Figure~\ref{fig:dynamictotal}(e). DG1
outperforms DFS starting with 2 queries per update.

In summary, over all the real and synthetic datasets, we observe
that typically DG1 gives the best overall performance, whereas
even DG0 (i.e., just maintaining the strongly connected
components without labeling) results in a significant improvement
over DFS. 
As expected, \alg amortizes the index maintenance cost against
query times, and thus the more the queries received, the more the
benefit.  It is also interesting to note that a single interval
label provides the best performance, and going to two labels does
not confer any effective advantage.  There are two reasons for
this: i) when the DAG corresponding to the input graph is very
sparse (or tree-like), a single interval is sufficient to provide
fast querying, since it captures most of the topology, and ii)
dynamically updating labels makes the labels less random, since
our current label propagation algorithm is deterministic, and of
course there is extra cost associated with label propagation.

%
{\small
\bibliographystyle{abbrv}
\bibliography{dagger}  
}
%
%
\end{document}